**Title:** PsSource: an installable Geant4 extension for transport-coupled positronium annihilation with validated Ore–Powell three-photon physics

**Authors:** Jason G Parker, PhD, Department of Radiology and Imaging Sciences, Indiana University School of Medicine, parkerjg@iu.edu

## Abstract

PsSource is an installable Geant4 physics extension for configurable positronium-aware terminal positron annihilation. In the recommended transport-coupled workflow, the host application creates the positron and Geant4 transports it through the existing geometry and materials. PsSource replaces only terminal at-rest annihilation, preserves the transported terminal position and global time, resolves a fixed or host-defined local positronium environment, samples the realized annihilation class and delay, and returns ordinary two- or three-photon Geant4 secondaries for continued transport. Supported outcomes are direct two-photon, para-positronium two-photon, ortho-positronium two-photon, and ortho-positronium three-photon annihilation. Fixed and exponential delay models and approximate, approximate phase-space, Geant4 Ore–Powell, and polarized Geant4 Ore–Powell three-photon backends are available. Optional observer and provenance interfaces support event-level recording without coupling the physics to PsSource-specific truth or output infrastructure.

Transport-coupled regression tests verified preservation of terminal position and time, decomposition of positron transport time and sampled positronium delay, recovery of all four annihilation classes, correct photon multiplicity and parentage, identical results for equivalent fixed and provider-returned environments, and unchanged photon physics with or without PsSource-specific truth recording. The PsSource integration of the native Geant4 Ore–Powell model was independently evaluated using 100,000 three-photon events. Comparison with an analytic joint-energy distribution yielded $\chi^2$ = 1236.91 for 1274 degrees of freedom, a standardized $\chi^2$ score of −0.735, and a maximum marginal empirical-CDF difference of 0.00208. Ordered photon-energy spectra also agreed with independent native Geant4 and GATE reference implementations, with no pairwise empirical-CDF difference greater than 0.00452. Photon directions and event-plane normals were consistent with rotational isotropy and showed no detectable Cartesian-axis bias. The polarized backend satisfied direction normalization, polarization normalization, and transversality to within $1.5\times10^{-12}$; the maximum empirical-CDF difference from the ordinary Ore–Powell spectrum was 0.00168.

These results establish PsSource as a reusable and validated integration layer for positronium-aware terminal annihilation in existing Geant4 applications. The current implementation does not model microscopic positronium formation, positronium-atom transport, collision-by-collision quenching, or automatic material-specific parameter inference.

## 1. Introduction

Positronium (Ps), the bound state of an electron and a positron, introduces annihilation-time, photon-multiplicity, angular-correlation, spin, and polarization information beyond the conventional representation of positron emission tomography as prompt emission of two nearly back-to-back 511-keV photons [1–6]. Para-positronium predominantly annihilates through the two-photon channel, whereas ortho-positronium intrinsically decays primarily through three-photon emission in vacuum [6,7]. In condensed matter and biological tissue, however, interactions with the surrounding medium substantially modify ortho-positronium behavior through two-photon pickoff annihilation and other quenching processes, often reducing the intrinsic three-photon contribution to a minor component [2–6]. The measured lifetimes and relative populations of direct annihilation, para-positronium, ortho-positronium two-photon annihilation, and ortho-positronium three-photon annihilation are therefore not universal properties determined by spin state alone, but depend on the local chemical, structural, and electronic environment [3–6].

Simulation of positronium-sensitive systems therefore requires more than assigning an annihilation delay or selecting a photon multiplicity. A simulation-relevant event description may include positron creation and transport, the terminal annihilation position and time, environment-dependent branch fractions and lifetime distributions, photon energies and directions, polarization states, and optional prompt-associated emission [3–6]. These quantities are not generally independent physical inputs. In real materials, branch fractions and lifetimes are coupled to the local environment, whereas in controlled simulation studies they may be prescribed deliberately to isolate detector, reconstruction, or analysis effects. Studies involving heterogeneous samples, multiple lifetime components, or spatially varying material properties further motivate interfaces in which local positronium parameters can be supplied independently of the underlying particle-transport and detector model [5, 8, 9]. Although PET is the principal application considered here, the same integration problem arises in positron-beam experiments, lifetime spectroscopy, accelerator-target studies, condensed-matter simulations, and other Geant4 applications involving transported positrons.

Geant4, GATE, and detector-specific research frameworks already provide important components of positronium and multiphoton-annihilation simulation. Geant4 supplies the underlying particle-transport and electromagnetic-physics infrastructure and, in recent releases, includes native positron-at-rest models implementing Ore–Powell three-photon annihilation with polarization support [10–12]. GATE extends Geant4 with source, digitization, and imaging-workflow tools and supports positronium lifetime modeling, mixed two- and three-photon emission, and prompt-photon configurations used in PET studies [13,14]. The J-PET collaboration has likewise developed specialized Geant4- and GATE-based approaches for positronium lifetime imaging, three-photon reconstruction, and polarization-sensitive measurements [2,3,15], while other direct Geant4 implementations have modeled positronium lifetime distributions for dedicated applications [16]. These established systems demonstrate that physically credible positronium and three-photon simulation components already exist. The contribution of PsSource is not to replace them, but to provide a process-level, installable interface that couples the transported Geant4 positron endpoint to host-defined positronium parameters and validated annihilation final states.

For an established custom Geant4 application, this is primarily an integration problem. Replacing a transported positron with a source-level photon generator bypasses its interaction history and fixes the annihilation state before Geant4 has determined the terminal position and time. A reusable extension should instead preserve the existing host simulation chain while acting only at the point where the transported positron undergoes terminal annihilation. The required computational sequence is therefore: the host application creates the positron, Geant4 transports it normally, a positronium-aware process resolves the local environment at the terminal state and samples the annihilation outcome, and the resulting photons return to ordinary Geant4 tracking. Separating local environmental parameterization from final-state generation also allows fixed, material-based, region-based, spatially varying, or future experimentally derived models to use the same interface.

PsSource implements this design as an installable Geant4 physics extension for configurable positronium-aware terminal positron annihilation. It replaces only the terminal at-rest annihilation process and preserves the transported positron position and global time. A fixed or host-defined environment provider supplies local branch fractions and lifetime parameters, from which PsSource samples one of four realized annihilation classes: direct two-photon, para-positronium two-photon, ortho-positronium two-photon, or ortho-positronium three-photon annihilation. The sampled positronium delay is added to the terminal transport time, and the resulting photons are created as ordinary Geant4 secondaries. Approximate phase-space, native Geant4 Ore–Powell, and native polarized Ore–Powell three-photon backends share the same process-level interface. Optional observer and provenance mechanisms support event-level recording without making the annihilation physics dependent on PsSource-specific truth or output infrastructure.

This terminal at-rest handoff is a computational integration boundary rather than a microscopic model of positronium formation. The current implementation does not model positronium formation during positron slowing, positronium-atom transport or diffusion, collision-by-collision pickoff or quenching, magnetic-substate-resolved dynamics, or automatic inference of material-specific positronium parameters. Environmental branch fractions and lifetimes remain application-supplied model inputs, and validation of the annihilation process does not establish the physical validity of a particular material or tissue parameterization.

In this work, we describe the transport-coupled PsSource architecture and evaluate both its Geant4 integration behavior and its annihilation final-state physics. Integration tests assess preservation of the transported terminal position and time, decomposition of positron transport time and sampled positronium delay, recovery of the four realized annihilation classes, photon multiplicity and parentage, identical results for equivalent fixed and provider-returned environments, and invariance of generated photon physics when PsSource-specific truth recording is disabled. The PsSource integration of the native Geant4 Ore–Powell models is evaluated separately against an independently calculated joint-energy distribution, native Geant4, and GATE, with additional tests of energy and momentum closure, rotational isotropy, Cartesian-axis symmetry, polarization normalization, direction–polarization transversality, event-plane polarization structure, and preservation of the Ore–Powell energy spectrum [7,12–15]. The objective is to establish PsSource as a reusable and validated integration layer for configurable positronium-aware terminal annihilation in existing Geant4 applications.

## 2. Methods

### 2.1 Software implementation and transport-coupled architecture

PsSource version 3.0 was implemented as an installable C++17 Geant4 physics extension for configurable positronium-aware terminal positron annihilation. The reusable library is exposed through the CMake target `PsSource::PsSource` and is integrated into a host application by registering `PsSourceAnnihilationPhysics` with an existing G4VModularPhysicsList. PsSource does not replace the host application's primary-particle source, geometry, materials, electromagnetic transport, detector construction, sensitive detectors, scoring, digitization, reconstruction, or output infrastructure.

In the transport-coupled workflow, the host application creates the positron and Geant4 transports it normally through the configured geometry and materials. When the positron reaches the terminal at-rest annihilation stage, PsSource replaces the existing Geant4 annihilation process, captures the transported terminal state, resolves the applicable positronium environment, samples the realized annihilation class and delay, generates the corresponding two- or three-photon final state, and returns the photons as ordinary Geant4 secondaries. Geant4 then transports those photons through the host application without requiring PsSource-specific downstream handling.

The architecture separates three responsibilities. First, the terminal-state integration layer captures the transported positron position, global time, kinetic energy, track identifiers, and available material and geometry context. Second, the environment and physics-model layer supplies the local direct-annihilation, para-positronium, and ortho-positronium fractions; lifetime parameters; ortho-positronium two- and three-photon branching; and the selected three-photon final-state model. Third, the Geant4 process layer terminates the positron and constructs the sampled annihilation photons at the preserved terminal position and at the terminal global time plus the sampled positronium delay.

The reusable public interface consists of a fixed or host-defined positronium environment, a configurable annihilation-physics constructor, an optional terminal-state environment provider, and an optional annihilation observer. When no environment provider is supplied, one fixed PsEnvironment parameterization is used for all annihilations. When a provider is supplied, the host application may resolve local parameters from the transported terminal state, including material, region, logical or physical volume, copy number, touchable hierarchy, position, or time. PsSource does not infer material- or tissue-specific positronium parameters from generic Geant4 material composition; the physical validity of the supplied parameterization remains the responsibility of the host application or of a separately validated environment model.

Annihilation observation is independent of the generated physics. The optional observer receives the realized positronium class, photon multiplicity, terminal positron time, sampled positronium delay, annihilation time and position, local environment identifier, lifetime, and model provenance. When no observer is configured, PsSource performs the same annihilation without producing PsSource-specific truth output. The reusable core therefore does not require `PositroniumTruthInfo`, a PsSource event or tracking action, a particular output schema, or the legacy explicit source-generation classes.

The current implementation supports direct two-photon annihilation, para-positronium two-photon annihilation, ortho-positronium two-photon annihilation, and ortho-positronium three-photon annihilation. Fixed and exponential delay models are available, together with approximate constrained phase-space, native Geant4 Ore–Powell, and native polarized Ore–Powell three-photon backends. The approximate backend is retained for controlled software testing and regression studies, whereas the Ore–Powell backends provide the principal physically validated three-photon final states used in this work. Model name, version, and validation status are retained as provenance independently of the selected host application or output format.

## 2.2 Terminal positron state and process replacement

PsSource acts only when a transported positron reaches the terminal Geant4 at-rest annihilation stage. Immediately before annihilation, the integration layer constructs a terminal-state description containing the source and terminal positions and times, initial and terminal kinetic energies, track and parent identifiers, track length, source-to-terminal displacement, and available geometry context, including the material, region, logical volume, physical volume, copy number, and touchable hierarchy. This information is passed to the environment and annihilation models without altering the preceding Geant4 transport history.

The annihilation position is defined by the transported terminal positron position, $r_{ann} = r_{e^+,terminal}$. No additional source-level positron-range displacement or positronium-atom displacement is applied in the transport-coupled pathway. The annihilation time is defined as $t_{ann} = t_{e+,terminal} + \Delta t_{Ps}$, where $t_{e+,terminal}$ is the transported positron global time and $\Delta t_{Ps}$ is the fixed or exponentially sampled positronium delay associated with the realized annihilation class. The generated photons therefore share the transported terminal position and the delayed annihilation time.

Process replacement is performed during Geant4 physics initialization. PsSource obtains the positron process manager, inspects the registered process list, and requires exactly one existing process named `annihil`. If no matching process or more than one matching process is found, initialization terminates with a fatal diagnostic rather than proceeding with an ambiguous annihilation configuration. The custom model replaces the terminal `AtRestDoIt` behavior used for positron annihilation at rest; it does not apply the configurable positronium final-state model to in-flight annihilation. This registration-level replacement, together with the fatal initialization checks, prevents native and PsSource terminal-annihilation mechanisms from being registered ambiguously or acting simultaneously.

When its terminal at-rest interaction is invoked, `PsSourceAnnihilationProcess` captures the terminal state, resolves the applicable local `PsEnvironment`, samples the positronium branch and delay, and generates the corresponding two- or three-photon final state. The transported positron is then terminated, and the annihilation photons are added as Geant4 secondary tracks with the positron as their parent. Their creator-process identity remains `annihil`, and subsequent photon transport, interaction, scoring, and detection are handled by the host application's existing Geant4 configuration.

The terminal at-rest handoff is a computational integration boundary rather than a microscopic model of positronium formation. PsSource does not assume that positronium physically forms only after the positron has reached rest, nor does the current implementation simulate formation during slowing, positronium diffusion or transport, or collision-by-collision pickoff and quenching. Instead, Geant4 models positron creation and transport up to the terminal state, after which PsSource applies the configured positronium-aware annihilation model.

## 2.3 Local positronium environment

Local positronium behavior is represented by a `PsEnvironment` structure that supplies the parameters used for one terminal annihilation. The environment includes an application-defined medium identifier; the relative fractions of direct annihilation, para-positronium, and ortho-positronium; the lifetime assigned to each class; and the relative two- and three-photon branching fractions for ortho-positronium. These quantities define the local parameterization used by PsSource but do not themselves constitute a microscopic material model.

PsSource supports two environment-resolution pathways. When no environment provider is configured, the fixed `PsEnvironment` stored in `PsSourceAnnihilationConfig` is applied to every terminal annihilation. This is the simplest integration mode and is appropriate for homogeneous simulations or controlled studies in which one parameter set is intentionally used throughout the geometry.

For spatially or materially varying applications, the host may supply an implementation of `IPsSourceEnvironmentProvider`. Immediately before branch and delay sampling, PsSource passes the transported `PsTerminalState` to the provider and receives the `PsEnvironment` applicable to that annihilation. The provider may resolve local parameters from the available terminal-state, geometry, tracking, or application-defined spatial data. This allows the host application to implement local lookup tables, spatial parameter maps, or externally derived environment models without modifying the PsSource annihilation process.

The class fractions are denoted $f_{direct}$, $f_{pPs}$, and $f_{oPs}$ and are required to be non-negative. The fractions are required to sum to unity within the configured numerical tolerance and define the probabilities of selecting direct annihilation, para-positronium, or ortho-positronium. For ortho-positronium events, the environment additionally supplies $f_{oPs,2\gamma}$ and $f_{oPs,3\gamma}$. These fractions are required to be non-negative and to sum to unity within the configured numerical tolerance, and they define the probabilities of the two- and three-photon final states. The associated lifetimes are denoted $\tau_{direct}$, $\tau_{pPs}$, and $\tau_{oPs}$. The same $\tau_{oPs}$ value is currently applied to both ortho-positronium final-state branches.

Environment resolution occurs at the transported terminal positron state rather than at the original source position. Consequently, two positrons created under identical source conditions may receive different parameterizations if they terminate in different local environments. The selected environment is then used consistently for class selection, delay sampling, final-state generation, and optional provenance reporting.

PsSource validates the numerical consistency of the returned environment before annihilation sampling. Invalid probability values, non-finite parameters, negative lifetimes, or inconsistent branch definitions are rejected rather than propagated into event generation. These checks establish numerical and internal consistency only; they do not validate the parameterization as representative of a specific material, tissue, or experimental condition.

PsSource does not infer positronium fractions or lifetimes automatically from generic Geant4 material composition, elemental content, density, or tissue name. Such relationships must be supplied by the host application or by a separately developed and validated environment model. The environment-provider interface therefore separates local material or biological parameterization from the annihilation-physics implementation and allows future experimentally derived models to use the same transport-coupled architecture.

**2.4 Branch selection and delay modeling**

For each terminal positron annihilation, PsSource first samples one of three positronium classes: direct annihilation, para-positronium, or ortho-positronium. The corresponding environment fractions are denoted $f_{direct}$, $f_{pPs}$, and $f_{oPs}$. They are required to be non-negative and to satisfy $f_{direct} + f_{pPs} + f_{oPs} = 1$. The selected class determines the lifetime parameter and the allowed annihilation topology.

Direct-annihilation and para-positronium events produce two-photon final states. Ortho-positronium events undergo a second branch decision between two- and three-photon annihilation according to $f_{oPs,2\gamma}$ and $f_{oPs,3\gamma}$, with $0 \le f_{oPs,2\gamma} \le 1, 0 \le f_{oPs,3\gamma} \le 1$, and $f_{oPs,2\gamma} + f_{oPs,3\gamma} = 1$. The four realized event classes are therefore direct $2\gamma$, p-Ps $2\gamma$, o-Ps $2\gamma$, and o-Ps $3\gamma$. The o-Ps $2\gamma$ branch provides a configurable representation of pickoff-dominated or otherwise quenched conditions, but PsSource does not simulate the microscopic collision processes that produce this branch in a specific material.

After the realized class is selected, PsSource samples the additional positronium delay, $\Delta t_{Ps}$. Two timing modes are supported. In fixed-delay mode, the single configured fixed delay is applied after selection of the realized class. In exponential mode, the delay is sampled as $\Delta t_{Ps} = -\tau \ln u$, where u is uniformly distributed on the open interval (0,1) and τ is the lifetime associated with the realized class. Direct, p-Ps, and o-Ps events use $\tau_{direct}$, $\tau_{pPs}$, and $\tau_{oPs}$, respectively. The same $\tau_{oPs}$ value is currently used for both o-Ps $2\gamma$ and o-Ps $3\gamma$ events. A zero value of $\tau_{direct}$ produces no additional delay for direct annihilation in exponential mode, whereas $\tau_{pPs}$ and $\tau_{oPs}$ must be strictly positive.

Uniform random variates are obtained from `G4UniformRand()`, which uses the Geant4-managed CLHEP random engine. Branch selection and delay sampling therefore remain within the host application's Geant4 random stream and are reproducible under fixed random seeds. No separate random-number generator is introduced by PsSource.

The sampled delay is added to the transported terminal positron global time rather than to the original source time. The final annihilation time is therefore $t_{ann} = t_{e+,terminal} + \Delta t_{Ps}$. This preserves the full positron transport time already accumulated by Geant4 and separates it from the additional delay assigned by the positronium model.

Configuration checks are performed before or during model initialization to reject invalid probabilities, non-finite values, negative lifetimes, and inconsistent branch fractions. These checks ensure internal numerical consistency but do not establish the physical validity of the selected branch fractions or lifetimes for a particular material.

**2.5 Annihilation final-state generation**

After the local environment has been resolved and the realized positronium class and delay have been sampled, PsSource generates the corresponding annihilation-photon final state. The PsSource orchestration layer returns the realized class, annihilation multiplicity, sampled positronium delay, photon final state, and model provenance. For three-photon events, the selected backend supplies the photon energies, propagation directions, and optional polarization vectors. The final-state generator does not modify the transported terminal positron position or time; it defines only the annihilation products created at that state.

Two-photon final states are used for direct annihilation, p-Ps annihilation, and the configured o-Ps $2\gamma$ branch. Three-photon final states are used only for the realized o-Ps $3\gamma$ class. For three-photon events, PsSource supports an approximate constrained phase-space backend, the native Geant4 Ore–Powell model, and the native polarized Ore–Powell model. Branch selection and delay sampling are common to all backends, so changing the three-photon model alters only the generated three-photon kinematics and polarization properties.

In the transport-coupled pathway, the generated photons are created as Geant4 secondaries of the transported positron at $r_{ann} = r_{e+,terminal}$ and $t_{ann} = t_{e+,terminal} + \Delta t_{Ps}$. Their energies, directions, and polarization states are transferred directly from the selected final-state model. The photons then continue through ordinary Geant4 tracking and interact with the host geometry and detector systems without PsSource-specific downstream processing.

The same physics-model interfaces are also used by the explicit benchmark application retained in the repository. In that pathway, isolated photon events are constructed for analytic validation, cross-framework comparison, and regression testing. The explicit pathway does not define the recommended integration architecture, but it allows the annihilation final-state distributions to be evaluated independently of positron transport and detector response.

2.5.1 Two-photon final states

Within the current at-rest final-state approximation, direct $2\gamma$, p-Ps $2\gamma$, and o-Ps $2\gamma$ events use the same two-photon kinematic construction. Each event produces exactly two photons with kinetic energies $E_1 = E_2 = m_e c^2$ and opposite unit propagation directions, $\widehat{k_2} = -\widehat{k_1}$. The first direction is sampled isotropically, and the second is defined by momentum conservation. The resulting final state therefore satisfies $E_1 + E_2 = 2m_e c^2$ and $\overrightarrow{p_1} + \overrightarrow{p_2} = 0$.

The three two-photon classes differ in their realized positronium identity and timing model rather than in their photon kinematics. Direct events use the direct-component lifetime or fixed delay, p-Ps events use the p-Ps lifetime or fixed delay, and o-Ps $2\gamma$ events use the o-Ps lifetime or fixed delay. The o-Ps $2\gamma$ class provides a configurable effective representation of pickoff or other quenched two-photon annihilation, but the final-state generator does not simulate the microscopic interaction responsible for that branch.

Both photons are created at the transported terminal positron position and at the same annihilation time. In transport-coupled operation, they are added as secondaries of the transported positron and retain the `annihil` creator-process identity. The current two-photon implementation does not assign photon polarization or model entangled polarization correlations; the generated photons are recorded as unpolarized unless a separate validated two-photon polarization model is introduced.

2.5.2 Approximate constrained phase-space backend

The approximate backend is retained for controlled software-development, sensitivity, and regression studies. It samples three positive photon energies subject to conservation of the total electron–positron rest energy,

$$E_1 + E_2 + E_3 = 2m_e c^2$$

with

$$0 < E_i < m_e c^2, \qquad i \in \{1,2,3\}$$

The corresponding photon momenta satisfy

$$p_1 + p_2 + p_3 = 0$$

with the photon relation

$$|p_i| = \frac{E_i}{c}$$

Emission directions are constructed in a randomly oriented event plane so that the three momentum vectors close exactly. The resulting events therefore satisfy photon multiplicity, positivity, energy conservation, and momentum conservation by construction.

These conditions are necessary but do not uniquely define the physical o-Ps 3γ distribution. In particular, this backend does not include the QED weighting of the Ore–Powell matrix element and is therefore designated in the model metadata as an approximate-controlled-source-model. It was not used as the principal physics backend for the Ore–Powell validation or cross-framework comparisons reported in this work.

### 2.5.3 Ore–Powell backend

For the ordinary Ore–Powell backend, PsSource delegates three-photon generation to the native Geant4 `G4OrePowellAtRestModel` through the PsSource physics-model interface. For each selected o-Ps 3γ event, the Geant4 model returns three secondary photons. PsSource extracts each photon's kinetic energy and momentum direction, verifies that exactly three photons were returned, and transfers these quantities into the common PsPhoton representation. Branch identity, lifetime, and event provenance remain under the PsSource orchestration layer. Although the underlying Geant4 model may populate polarization values, PsSource treats this backend as unpolarized and does not expose those vectors as validated polarization output.

Photon energies are expressed using the dimensionless variables

$$r_i = \frac{E_i}{m_e c^2}$$

which satisfy

$$r_1 + r_2 + r_3 = 2$$

and

$$0 < r_i < 1$$

for physical three-photon annihilation at rest. The pairwise opening angles are fixed by energy–momentum conservation. For photons i, $j$, and $k$,

$$\cos \theta_{ij} = \frac{r_k^2 - r_i^2 - r_j^2}{2 r_i r_j}$$

where $i$, $j$, and $k$, are distinct. Unlike the approximate backend, the probability assigned across the allowed phase space follows the QED-weighted Ore–Powell distribution [7]. PsSource delegates three-photon generation to the native Geant4 11.3.2 Ore–Powell model and is identified in the PsSource metadata as Geant4OrePowell, with validation status geant4-native-ore-powell [12].

### 2.5.4 Polarized Ore–Powell backend

The polarized backend uses the native Geant4 G4PolarizedOrePowellAtRestModel. It produces the same three-photon Ore–Powell annihilation topology while additionally assigning a polarization vector to each photon. PsSource extracts the polarization from the Geant4-generated photon, stores it in the corresponding `PsPhoton`, and sets the polarization-valid flag. In the explicit benchmark pathway, the photons are constructed as Geant4 primaries; in transport-coupled operation, they are constructed as Geant4 secondaries with the polarization assigned to the corresponding photon object.

For each photon, the propagation direction and polarization vector are denoted by

$$\vec{k_i}$$

and

$$\varepsilon_i$$

respectively. Valid photon states are required to satisfy unit normalization,

$$|\widehat{k_i}| = 1$$

$$|\varepsilon_i| = 1$$

and transversality,

$$\widehat{k_i} \cdot \varepsilon_i = 0$$

The relationship between polarization and the three-photon decay plane was evaluated using the unit event-plane normal

$$\hat{n} = \frac{k_1 \times k_2}{|k_1 \times k_2|}$$

and the absolute out-of-plane projection

$$|\varepsilon_i \cdot \hat{n}|$$

The corresponding in-plane polarization-component magnitude was computed as

$$\sqrt{1 - (\varepsilon_i \cdot \hat{n})^2}$$

The polarized backend is recorded as Geant4PolarizedOrePowell, with version metadata corresponding to Geant4 11.3.2 and validation status geant4-native-ore-powell [12]. The present implementation records photon polarization vectors but does not record the o-Ps magnetic substate. Accordingly, the validation performed here addresses vector normalization, transversality, event-plane structure, global-axis isotropy, and preservation of the Ore–Powell energy distribution, but not magnetic-substate-resolved observables.

## 2.6 Geant4 secondary construction and optional observation

After the positronium class, delay, and annihilation final state have been sampled, `PsSourceAnnihilationProcess` converts the model output into Geant4 secondary particles. All annihilation photons are created at the transported terminal positron position, $r_{ann} = r_{e+,terminal}$, and at the annihilation time, $t_{ann} = t_{e+,terminal} + \Delta t_{Ps}$. The final-state model supplies each photon's kinetic energy, propagation direction, and, when available, polarization vector.

Each generated photon is represented as a `G4DynamicParticle` of type `G4Gamma` and is added to the Geant4 particle-change object as a secondary of the transported positron. The photon directions are required to be normalized, the kinetic energies must be positive, and polarized photons must contain valid polarization vectors transverse to their propagation directions. The transported positron is then terminated, and the annihilation final state carries the modeled electron–positron rest energy rather than being generated by the native terminal-annihilation process.

The generated photons inherit the transported positron as their parent track and retain `annihil` as their creator-process identity. Their birth position and time are determined by the terminal positron state and sampled positronium delay, not by the original primary-particle vertex. Once created, the photons are ordinary Geant4 secondaries and are transported by the host application's existing electromagnetic physics, geometry, fields, sensitive detectors, scoring, and digitization systems. PsSource does not require specialized photon tracking or detector-response code.

Event observation is optional and is separated from final-state generation through the `IPsSourceAnnihilationObserver` interface. When an observer is configured, PsSource supplies one `PsSourceAnnihilationRecord` for each realized annihilation. The record contains the realized positronium class, two- or three-photon multiplicity, transported terminal positron time, sampled positronium delay, final annihilation time and position, application-defined medium identifier, lifetime used for the realized class, and the selected model name, version, and validation status.

The observer receives metadata after the annihilation has been realized but does not own, resample, or modify the generated photons. The host application may translate the record into its existing output system, including CSV, ROOT, HDF5, JSON, or application-specific event structures. Photon-level information may be obtained through ordinary Geant4 tracking or scoring when required; it is not currently included in the public annihilation record.

When no observer is supplied, PsSource performs the same branch selection, delay sampling, and photon generation without producing PsSource-specific annihilation records. The absence of observation therefore does not alter photon multiplicity, energy, direction, polarization, position, time, or parentage. This design allows PsSource to operate in applications with no custom event information, event action, tracking action, or PsSource-specific output format.

### 2.7 Explicit source-generation and benchmark pathway

PsSource retains an explicit source-generation pathway for isolated validation and controlled software studies. In this mode, positronium branch selection, delay sampling, and annihilation final-state generation are performed without relying on a transported terminal positron. The resulting photons are converted into explicit Geant4 primary vertices rather than being created as secondaries of a transported positron.

This pathway is used for the analytic Ore–Powell benchmark, native Geant4 and GATE comparisons, polarization studies, isotropy testing, and backend regression. Isolating the annihilation final state removes positron transport, detector acceptance, energy resolution, and reconstruction effects, allowing the generated photon distributions to be evaluated directly.

Both explicit and transport-coupled pathways use the same branch, delay, and three-photon model interfaces. Consequently, the explicit benchmarks test the same annihilation final-state backends used by the reusable transport-coupled extension, while the transport-coupled regression suite separately verifies terminal-state preservation, secondary construction, parentage, and timing integration.

### 2.8 Transport-coupled integration validation

Transport-coupled validation was performed separately from the isolated final-state benchmarks to verify that PsSource operated correctly within ordinary Geant4 positron transport. The regression suite tested preservation of the transported terminal state, correct realization of the configured positronium class, delay propagation, secondary-photon construction, parentage, process provenance, environment resolution, and independence from optional truth recording.

Deterministic tests were performed for four pure annihilation configurations: direct $2\gamma$ with zero additional delay, p-Ps $2\gamma$ with a fixed delay, o-Ps $2\gamma$ with a fixed delay, and o-Ps $3\gamma$ with a fixed delay. For each configuration, the requested class fraction was set to unity and all competing branches were disabled. Each event was required to contain one terminal annihilation, the expected realized class, the expected two- or three-photon multiplicity, and no prompt-associated photon.

For every annihilation, the photon birth position was compared with the transported terminal positron position. Each deterministic configuration used 100 events. The position residual was defined as $\delta r = |r_\gamma - r_{e+,terminal}|$, where $r_\gamma$ is the common photon creation position. The timing residual was defined as $\delta t = t_\gamma - t_{e+,terminal} - \Delta t_{Ps}$, where $t_\gamma$ is the photon creation time and $\Delta t_{Ps}$ is the sampled positronium delay. All photons from a given annihilation were required to share their creation position and time to within $10^{-9}$ mm and $10^{-9}$ ns, respectively. Agreement between the photon and annihilation-summary positions and times, including the timing decomposition, was required to within $10^{-6}$ mm and $10^{-6}$ ns.

Secondary-track checks verified that the annihilation photons were created by the `annihil` process, inherited the transported positron as their parent, and received distinct Geant4 track identifiers. For two-photon events, each photon energy was required to agree with $m_ec^2$ within $10^{-9}$ MeV, each direction norm was required to agree with unity within $10^{-9}$, and the norm of the summed direction vectors was required to be below $10^{-9}$. For three-photon events, the regression suite required three photons with positive energies, direction norms within $10^{-9}$ of unity, total energy within $10^{-9}$ MeV of $2m_ec^2$, and momentum-closure residual below $10^{-9}$ MeV/c. These conservation checks were treated as implementation tests and not as validation of the physical three-photon distribution.

Exponential-delay behavior was tested independently for pure p-Ps $2\gamma$ and o-Ps $3\gamma$ configurations. Each exponential-delay configuration used 10,000 events. For each case, the sampled delays were compared with the expected cumulative distribution $F(t) = 1 - \exp\left(-\frac{t}{\tau}\right)$, where $\tau$ was the lifetime assigned to the realized class. Agreement was evaluated using the one-sample Kolmogorov–Smirnov statistic, together with the observed sample mean and the exact timing decomposition $t_{ann} = t_{e+,terminal} + \Delta t_{Ps}$. For $N = 10{,}000$, the predefined acceptance limit was $D < \frac{1.63}{\sqrt{N}} = 0.0163$.

Environment-provider behavior was tested by comparing two otherwise identical deterministic runs: one using the fixed `PsEnvironment` stored directly in the annihilation configuration and one using `FixedPsSourceEnvironmentProvider` to return the same parameters. Under identical random-engine conditions, the annihilation summaries and photon-level outputs were required to be byte-for-byte identical. This deterministic software-equivalence test verified that the fixed-configuration and fixed-provider pathways produced identical outputs when they returned the same environment under identical random-engine conditions. It was not intended to establish general equivalence among arbitrary provider implementations.

The optional-observation design was evaluated by repeating the direct $2\gamma$ zero-delay case with PsSource-specific truth recording enabled and disabled. Photon-level outputs were compared for event identity, track and parent identifiers, creator process, birth position, birth time, kinetic energy, direction, polarization state, and polarization-valid flag. The generated photon physics was required to be identical between the two runs, demonstrating that truth collection and observer infrastructure do not participate in branch selection, delay sampling, or final-state generation.

Each regression case returned an explicit `PASS` or `FAIL` result. A case passed only when the requested number of events was recovered, all class and multiplicity checks succeeded, position and timing residuals remained within tolerance, secondary parentage and process provenance were correct, and the expected deterministic or statistical timing behavior was observed.

**2.9 Independent analytic Ore–Powell validation**

The PsSource Ore–Powell backend was validated against an independently evaluated analytic reference distribution rather than against conservation constraints alone. The comparison used 100,000 generated o-Ps 3γ events, corresponding to 300,000 annihilation photons. For each event, the photon energies were normalized as

$$r_i = \frac{E_i}{m_e c^2}$$

with

$$r_1 + r_2 + r_3 = 2$$

and

$$0 < r_i < 1$$

The analytic Ore–Powell density was evaluated over the physically allowed two-dimensional energy phase space using the ordinary unpolarized three-photon annihilation distribution [7]. The numerical reference grid used 50 bins per energy axis. Expected probabilities were obtained by numerical integration of the analytic density within each bin using 12 subdivisions per axis. The resulting expected probability in bin $j$, denoted $p_j$, was converted to an expected count according to

$$E_j = N p_j$$

where $N$ is the number of analyzed events.

The PsSource sample was reduced to the same two-dimensional energy representation and binned on the identical grid. For each populated comparison bin, the standardized residual was calculated as

$$R_j = \frac{O_j - E_j}{\sqrt{E_j}}$$

where $O_j$ and $E_j$ are the observed and expected counts, respectively. The global agreement statistic was calculated as

$$\chi^2 = \sum_j \frac{\left(O_j - E_j\right)^2}{E_j}$$

using only bins that satisfied the expected-count inclusion criterion. Bins were included when $E_J \geq 5$. No parameters were fitted to the PsSource sample; the analytic distribution, normalization, grid, and inclusion threshold were specified independently. The number of degrees of freedom was the number of included bins minus one because the expected and observed counts were constrained to the same fixed total event count; no additional degrees of freedom were subtracted for fitted parameters. To express the global result relative to the expected chi-square distribution, a standardized chi-square score was also computed as

$$z_{\chi^2} = \frac{\chi^2 - \nu}{\sqrt{2\nu}}$$

where $\nu$ is the number of degrees of freedom.

One-dimensional marginal distributions were obtained from the analytic reference and the sampled PsSource events. Agreement was quantified using the maximum absolute difference between the corresponding cumulative-distribution functions,

$$D_{max} = \max_r \left| F_{\text{sample}}(r) - F_{\text{analytic}}(r) \right|$$

The observed value was compared with a predefined tolerance of 0.01897. Additional event-level checks were performed for photon multiplicity, physical phase-space membership, direction-vector normalization, total normalized energy, and pairwise opening-angle consistency. Pairwise opening angles were compared with the values implied by the sampled photon energies,

$$\cos\theta_{ij} = \frac{r_k^2 - r_i^2 - r_j^2}{2 r_i r_j}$$

where $i$, $j$, and $k$ are distinct photon indices.

The validation was considered successful only if all multiplicity and phase-space checks passed, the maximum marginal CDF difference remained below tolerance, the global chi-square result was consistent with statistical sampling, and the direction, energy-sum, and opening-angle errors remained within their numerical tolerances. The corresponding density, sampled distribution, and standardized-residual map are shown in **Figure 1**.

### 2.10 Native Geant4 and GATE reference implementations

Reference datasets were generated through native Geant4 and GATE execution pathways to determine whether the PsSource Ore–Powell backend reproduced established three-photon annihilation implementations. The comparison was restricted to source-level photon energies from o-Ps $3\gamma$ annihilation at rest. Prompt-photon emission and positron-range displacement were disabled, and detector transport, energy resolution, acceptance, and reconstruction effects were excluded. The comparison therefore tested the generated annihilation distributions rather than downstream detector response.

The native Geant4 reference and the PsSource Ore–Powell backend use the same underlying Geant4 final-state model. Their comparison therefore tests process integration, event construction, photon extraction, and data reduction rather than providing independent validation of the Ore–Powell physics. GATE provides a separate source and execution pathway but is also built on Geant4. Independent validation of the physical energy distribution was provided by the analytic comparison in Section 2.9.

### 2.10.1 Native Geant4 reference

The native reference was generated with Geant4 version 11.3.2 using the FTFP_BERT reference physics list with `G4EmLivermorePolarizedPhysics`. Positron annihilation at rest was configured through the Geant4 electromagnetic interface using the `OrePowellPolar` model. In this pathway, a near-rest positron was generated and the subsequent annihilation was handled by native Geant4 physics rather than by explicit PsSource photon-vertex construction. The world material was `G4_AIR`, prompt-associated emission was disabled, and the native material-side ortho-positronium fraction was configured to produce the $3\gamma$ reference sample. Geant4 provides the transport and electromagnetic-physics infrastructure used by both the native and PsSource pathways [10-12]. Although the native reference used the polarized Ore–Powell model, only photon energies were retained for this comparison; polarization information was not analyzed, and the ordinary and polarized Ore–Powell backends were evaluated separately for energy-spectrum consistency.

The energies of the three annihilation photons were recovered from the native event output. Events were retained only when exactly three annihilation photons were identified. No photon-energy or angular acceptance threshold was applied for the energy-spectrum comparison. For each accepted event, the photon energies were normalized as

$$r_i = \frac{E_i}{m_e c^2}$$

and sorted in ascending order,

$$r_{(1)} \leq r_{(2)} \leq r_{(3)}$$

where $r_{(1)}$, $r_{(2)}$, and $r_{(3)}$ denote the lowest-, middle-, and highest-energy photons, respectively. This ordering removed arbitrary photon labels and allowed direct comparison of corresponding marginal distributions across implementations.

### 2.10.2 GATE reference

The GATE reference dataset was generated with GATE version 9.4.1, built against Geant4 version 11.3.0, using the GATE Extended source configured as an ortho-positronium point source. The GATE configuration selected o-Ps 3γ annihilation and disabled prompt-associated emission and detector-response effects so that the emitted annihilation-photon energies could be compared directly with the PsSource and native Geant4 samples. The simulation was executed in an Apptainer 1.3.6 container built from opengatecollaboration/gate:latest on March 17, 2026. GATE is built on Geant4 and provides source, particle-transport, digitization, and PET workflow functionality, including positronium and mixed 2γ/3γ source configurations [13, 14].

The emitted GATE photons were grouped by annihilation event, and only events containing exactly three annihilation photons were retained. A total of 100,168 events containing exactly three annihilation photons were retained, corresponding to 300,504 photons. Photon energies were converted to the same normalized variable,

$$r_i = \frac{E_i}{m_e c^2}$$

and ordered identically as

$$r_{(1)} \leq r_{(2)} \leq r_{(3)}$$

before construction of the marginal probability-density and empirical cumulative-distribution functions. No detector smearing, energy window, or reconstruction selection was applied.

2.10.3 Harmonization of the comparison datasets

The PsSource comparison sample was generated with the explicit Ore–Powell backend, with

$$f_{\text{direct}} = 0$$
$$f_{\text{pPs}} = 0$$
$$f_{\text{oPs}} = 1$$

and

$$f_{\text{oPs},3\gamma} = 1$$

Prompt-associated emission and positron-range displacement were disabled. The native Geant4 and GATE samples were configured to represent the corresponding o-Ps 3γ source condition. All three datasets were reduced to the same event-level representation: three normalized photon energies ordered from lowest to highest. This harmonization avoided comparisons based on implementation-specific particle ordering or output conventions.

The three marginal energy distributions were compared using normalized histograms and empirical cumulative-distribution functions. The cross-framework analysis and associated statistics are described in Section 2.11.

## 2.11 Cross-framework energy-spectrum comparison

Agreement among PsSource, native Geant4, and GATE was evaluated using source-level o-Ps 3γ photon-energy distributions. The PsSource and native Geant4 datasets each contained 100,000 accepted three-photon events, whereas the GATE dataset contained 100,168 accepted three-photon events reconstructed from the recorded spherical phase-space output. No detector response, energy resolution, acceptance window, or reconstruction filter was applied.

For each accepted event, the three photon energies were normalized as

$$r_i = \frac{E_i}{m_e c^2}$$

and sorted in ascending order,

$$r_{(1)} \leq r_{(2)} \leq r_{(3)}$$

where $r_{(1)}$, $r_{(2)}$, and $r_{(3)}$ represent the lowest-, middle-, and highest-energy photons, respectively. Sorting removed implementation-dependent photon labels and permitted direct comparison of corresponding marginal distributions.

Normalized probability-density histograms were constructed separately for the three ordered photon ranks. The same bin edges and normalization procedure were applied to all frameworks. Visual agreement was assessed from the superimposed distributions, and quantitative agreement was evaluated using empirical cumulative-distribution functions.

For each ordered photon rank, the empirical CDF was defined as

$$F_n(r) = \frac{1}{n}\sum_{j=1}^{n} I(r_j \leq r)$$

where $I$ is the indicator function and $n$ is the number of accepted events. Pairwise CDF differences were calculated for PsSource versus native Geant4, PsSource versus GATE, and native Geant4 versus GATE:

$$\Delta F_{A-B}(r) = F_A(r) - F_B(r)$$

The maximum absolute pairwise difference was then computed as

$$D_{max}^{A,B} = \max_{r}|F_A(r) - F_B(r)|$$

This statistic was evaluated independently for the lowest-, middle-, and highest-energy photon distributions. Because the samples were generated independently and were not event matched, the comparison was interpreted as a distributional consistency test rather than a photon-by-photon equivalence test.

The cross-framework analysis was treated as a descriptive comparison of implementation-level agreement rather than as a formal hypothesis test. Agreement was summarized by the superimposed ordered-energy spectra and the reported pairwise empirical-CDF differences; no separate acceptance threshold was applied. The resulting marginal distributions and pairwise CDF-difference curves are shown in **Figure 2**.

**2.12 Rotational isotropy and coordinate-axis-bias testing**

Rotational isotropy was evaluated for both individual photon directions and the orientation of the three-photon event plane. The analysis used 100,000 o-Ps 3γ events generated with the PsSource Ore–Powell backend, corresponding to 300,000 annihilation photons. Photon directions were represented by unit vectors

$$\widehat{k_i} = (k_{x,i}, k_{y,i}, k_{z,i})$$

For each photon, the polar and azimuthal coordinates were calculated as

$$\cos\theta_i = k_{z,i}$$

and

$$\phi_i = \text{atan2}(k_{y,i}, k_{x,i})$$

with

$$-1 \leq \cos\theta_i \leq 1$$

and

$$-\pi \leq \phi_i < \pi$$

For an isotropic distribution, $cos\theta$ is uniform on $[-1,1]$ and $\phi$ is uniform on $[-\pi, \pi)$. The sampled photon distributions were compared with these expectations using fixed-width histograms and chi-square goodness-of-fit tests.

The three-photon event-plane orientation was evaluated independently. For each event, the unit normal to the decay plane was constructed from two non-collinear photon directions:

$$\hat{n} = \frac{k_1 \times k_2}{|k_1 \times k_2|}$$

To represent each unoriented plane once, the normal was reflected when necessary so that $n_z \geq 0$. Under this positive-hemisphere convention, the plane inclination was represented by $\cos\theta_n = n_z$, which is uniformly distributed on $[0,1]$ for isotropically oriented unoriented planes. The event-plane azimuth was then calculated from the sign-oriented normal as $\phi_n = \mathrm{atan2}(n_y, n_x)$ and mapped to $[0,2\pi)$.

The event-plane-normal distributions were tested against the same isotropic expectations used for the photon directions. Events with numerically degenerate cross products were excluded only if the normal could not be defined within floating-point tolerance.

To test directly for Cartesian-axis preference, the absolute directional components were averaged over all annihilation photons:

$$\langle |k_x| \rangle$$
$$\langle |k_y| \rangle$$
$$\langle |k_z| \rangle$$

For an isotropic unit-vector distribution, each quantity has the expectation

$$\langle |k_x| \rangle = \langle |k_y| \rangle = \langle |k_z| \rangle = \frac{1}{2}$$

The maximum pairwise Cartesian difference was computed as

$$\Delta_{\mathrm{axis}} = \max\left(\left|\langle |k_x| \rangle - \langle |k_y| \rangle\right|, \left|\langle |k_x| \rangle - \langle |k_z| \rangle\right|, \left|\langle |k_y| \rangle - \langle |k_z| \rangle\right|\right)$$

This metric was used as a direct test for preferred x-, y-, or z-axis alignment.

The angular tests were considered successful when the photon-direction and event-plane-normal distributions were statistically consistent with uniformity and the Cartesian-component means remained consistent with the isotropic expectation. The corresponding distributions and summary metrics are shown in **Figure 3**.

### 2.13 Polarization validation

Polarization validation was performed using 100,000 o-Ps 3γ events generated with the polarized Ore–Powell backend, corresponding to 300,000 annihilation photons. The analysis tested four properties: normalization and transversality of the polarization vectors, their relationship to the three-photon decay plane, absence of global-axis bias, and preservation of the underlying Ore–Powell photon-energy distribution.

#### 2.13.1 Vector normalization and transversality

For each photon, the unit propagation direction and polarization vector were denoted by $k_i$ and $\varepsilon_i$, respectively. The direction-norm error was calculated as

$$\delta_{k,i} = \left| |k_i| - 1 \right|$$

and the polarization-norm error as

$$\delta_{\varepsilon,i} = \left| |\varepsilon_i| - 1 \right|$$

Transversality was evaluated from the absolute direction–polarization scalar product,

$$\delta_{\perp,i} = |k_i \cdot \varepsilon_i|$$

For each metric, the maximum value over all photons was retained:

$$\delta_k^{max} = \max_i(\delta_{k,i})$$

$$\delta_\varepsilon^{max} = \max_i(\delta_{\varepsilon,i})$$

$$\delta_\perp^{max} = \max_i(\delta_{\perp,i})$$

The validation tolerance for direction normalization, polarization normalization, and transversality was

$$\delta_{tol} = 2 \times 10^{-5}$$

This was a predefined numerical-validity threshold for detecting malformed vectors or failed polarization propagation; it was not an estimate of the expected floating-point precision. The vector checks passed only if all three maximum errors remained below this threshold. Photon polarization validity was additionally required for every photon in the polarized sample. The corresponding validation routines were applied directly to the photon directions and polarization vectors retained in the event-level output.

### 2.13.2 Polarization relative to the decay plane

For each three-photon event, the unit normal to the decay plane was calculated from two non-collinear photon directions:

$$\hat{n} = \frac{k_1 \times k_2}{|k_1 \times k_2|}$$

The absolute polarization projection normal to the event plane was then calculated as

$$P_{\perp,i} = |\varepsilon_i \cdot \hat{n}|$$

Because each polarization vector was normalized, the magnitude of its in-plane component was calculated as

$$P_{\parallel,i} = \sqrt{1 - (\varepsilon_i \cdot \hat{n})^2}$$

Distributions of $P_\perp$ and $P_\parallel$ were generated over all annihilation photons. Their shapes and sample means were used to characterize the relationship between polarization and the three-photon decay plane. This analysis was intended to verify nontrivial decay-plane-dependent polarization structure rather than only vector normalization.

### 2.13.3 Global-axis isotropy

Potential polarization bias relative to the simulation coordinate system was evaluated using the absolute projection onto the global z axis:

$$P_{z,i} = |\varepsilon_i \cdot \hat{z}|$$

For an isotropically oriented population of unit polarization vectors, $P_z$ is uniformly distributed on $[0,1]$, with expectation

$$\langle P_z \rangle = \frac{1}{2}$$

The sampled distribution of the absolute global-$z$ projection, $|\varepsilon \cdot \hat{z}|$ was compared with this uniform expectation using fixed-width histograms and an angular-uniformity test. The goodness-of-fit statistic was calculated as a chi-square value over 50 histogram bins, with 49 degrees of freedom. A standardized chi-square score was computed as

$$z_{\chi^2} = \frac{\chi^2 - \nu}{\sqrt{2\nu}}$$

where $\nu$ is the number of degrees of freedom.

The sample mean was also compared directly with 0.5. This analysis tested whether the polarization-generation procedure introduced a preferred global coordinate axis.

2.13.4 Preservation of the Ore–Powell energy distribution

The polarized backend was tested to determine whether enabling polarization altered the ordinary Ore–Powell photon-energy distribution. Independent polarized and ordinary Ore–Powell samples were generated under otherwise identical o-Ps $3\gamma$ conditions and compared using their empirical photon-energy cumulative-distribution functions. The test statistic was the maximum absolute CDF difference,

$$D_{max}^{\text{pol-unpol}} = \max_{r} \left| F_{\text{pol}}(r) - F_{\text{unpol}}(r) \right|$$

The polarized backend passed this test when $D_{max}^{\text{pol-unpol}} < 0.020$. The threshold of 0.020 was specified in the validation workflow before evaluation as a conservative software-regression criterion for the independently generated 100,000-event samples. It was not interpreted as a formal two-sample hypothesis-test critical value.

2.13.5 Scope of the polarization validation

The current truth schema records one polarization vector for each generated photon but does not record the magnetic substate of the parent o-Ps system. The present analysis therefore validates photon-level normalization, transversality, decay-plane dependence, global-axis isotropy, and energy-spectrum preservation. It does not evaluate magnetic-substate-resolved polarization observables. The polarization results are summarized in **Figure 4**.

**2.14 Software build, installation, and verification**

Software verification was organized into build and installation tests, transport-coupled integration regressions, and annihilation-physics regressions. These automated tests were used to verify software behavior and reproducibility; the scientific validation methods for transport timing, Ore–Powell distributions, geometry, isotropy, and polarization are described separately in Sections 2.8–2.13.

PsSource was built as a C++17 CMake project against Geant4 11.3.2. The build system produces the reusable PsSource library, the namespaced target `PsSource::PsSource`, installed public headers, and CMake package-configuration files supporting downstream discovery with `find_package(PsSource REQUIRED)`. A clean parallel build was required to complete without compiler or linker errors.

CTest software-integration tests exercised three standalone host applications linked against the reusable library. The standalone transport test verified registration of `PsSourceAnnihilationPhysics` within an existing Geant4 physics list and successful generation of transport-coupled annihilation photons. The fixed-environment-provider test verified the public provider interface using one constant `PsEnvironment`. The local-environment-resolution test verified that a host-defined provider could select different annihilation parameters from terminal-state geometry context. Each test was executed through CTest and required an explicit passing result.

A separate smoke-build workflow tested both in-tree and installed-package use. PsSource was installed to a temporary prefix, after which an independent downstream application was configured and built using `find_package(PsSource REQUIRED)` and linked only through `PsSource::PsSource`. This test verified installation of the public headers, exported CMake targets, package configuration, transitive Geant4 dependencies, and the intended external integration pattern.

The transport-coupled regression suite automated the integration tests described in Section 2.8. It covered deterministic direct $2\gamma$, p-Ps $2\gamma$, o-Ps $2\gamma$, and o-Ps $3\gamma$ configurations; exponential p-Ps and o-Ps delays; equivalence of fixed configuration and fixed-provider resolution; and invariance of photon physics when PsSource-specific truth recording was disabled. Each case was validated from annihilation-summary and photon-level outputs and returned an explicit `PASS` or `FAIL` status.

A separate physics regression suite exercised the approximate constrained phase-space, Ore–Powell, and polarized Ore–Powell three-photon backends under standardized pure o-Ps $3\gamma$ configurations. The suite verified backend selection, event and photon multiplicity, model name, model version, validation-status provenance, directional isotropy, and the expected presence or absence of valid polarization information. Polarized events were additionally required to satisfy the vector-validity checks described in Section 2.13. Invalid command-line model names were required to fail with a diagnostic message rather than silently selecting another backend.

The physics regression workflow also generated standardized output from all three backends and performed an automated distributional comparison. This comparison was used to detect unintended changes to the generated energy spectra and to distinguish the approximate phase-space backend from the two native Ore–Powell pathways. It supplemented, but did not replace, the independent analytic and cross-framework validation described in Sections 2.9–2.11.

All build, installation, CTest, transport-regression, and physics-regression checks were required to pass before the PsSource 3.0.0 release used in this study. Geant4 11.3.2 was the validated release baseline; broad compatibility across other Geant4 versions was not established in this study. The associated build scripts, regression drivers, validators, test applications, and machine-readable outputs are included in the public repository.

**2.15 Software and data availability**

PsSource version 3.0.0 was used for this study. The corresponding public release is identified by the Git tag pssource-v3.0.0 and release commit 7dd7fdc. The software is available at https://github.com/parkerjg/PsSource under the MIT license. The immutable pssource-v3.0.0 release tag, rather than the moving main branch, is the authoritative software version for this manuscript.

PsSource is distributed as an installable C++17 CMake package. Downstream applications may locate the package with `find_package(PsSource REQUIRED)` and link against the exported target `PsSource::PsSource`. The installed public interface includes the annihilation-physics constructor, configuration and type definitions, terminal-state representation, fixed and host-defined environment-provider interfaces, and optional annihilation-observer interface. The reusable library does not require the repository's legacy explicit source generator, application-specific truth classes, or validation executables.

The principal validation baseline was Geant4 11.3.2. Public regression scripts and standalone test applications reproduce the build, installation, transport-coupled integration, environment-provider, truth-independence, and three-photon backend tests described in this work. The repository also includes the benchmark configurations, validation scripts, and processed data used to generate the analytic Ore–Powell, cross-framework, geometry, isotropy, and polarization results.

The native Geant4 reference was generated with Geant4 11.3.2. The independent GATE reference was generated with GATE 9.4.1 built against Geant4 11.3.0 using the documented containerized configuration. The repository records the software versions, random seeds, run configurations, and processing procedures used for these reference datasets.

Source code, build and installation instructions, regression drivers, benchmark definitions, native Geant4 and GATE reference configurations, and figure-ready processed data are provided in the public repository. Detailed artifact names, exact execution commands, and output-file descriptions are documented in the repository README and validation directories rather than repeated here.

## 3. Results

PsSource was evaluated in two complementary settings. Transport-coupled regressions tested preservation of the Geant4 positron terminal state, realized annihilation class, photon multiplicity and parentage, delay propagation, environment-provider behavior, and independence from optional PsSource truth recording. Separate isolated final-state benchmarks evaluated the Ore–Powell energy distribution, agreement with native Geant4 and GATE, three-photon geometry and rotational isotropy, and polarization behavior. All PsSource build, installation, integration, and physics-regression workflows completed successfully for the PsSource 3.0.0 release using the Geant4 11.3.2 validation baseline.

### 3.1 Transport-coupled integration

In all transport-coupled regression cases, Geant4 transported the primary positron from its source position to a terminal at-rest state before PsSource generated the annihilation final state. The resulting source-to-terminal displacement was nonzero and was therefore determined by the configured Geant4 transport and material environment rather than by a PsSource source-level displacement model.

For each annihilation, the generated photons originated at the transported terminal positron position. The terminal-to-annihilation position residual,

$$| \, r_{ann} - r_{e+,terminal} \, |,$$

was no greater than $1.0\times10^{-6}$ mm across the deterministic and exponential-delay cases. The corresponding timing residual,

$$t_{ann} - t_{e+,terminal} - \Delta t_{Ps},$$

was no greater than $1.0\times10^{-6}$ ns, confirming that the annihilation time was formed from the accumulated Geant4 transport time plus the sampled positronium delay.

The four deterministic configurations each recovered the expected realized annihilation class and photon topology in all 100 events. Direct annihilation produced class 0 with two photons per event and zero added delay; p-Ps produced class 1 with two photons per event and a fixed 3-ns delay; o-Ps two-photon annihilation produced class 2 with two photons per event and a fixed 3-ns delay; and o-Ps three-photon annihilation produced class 3 with three photons per event and a fixed 3-ns delay. All generated photons were ordinary Geant4 secondaries with the transported positron as parent and `annihil` as the creator process.

The exponential-delay regressions each included 10,000 events. The empirical p-Ps delay distribution agreed with the configured exponential model with $\tau = 0.125$ ns, yielding a Kolmogorov–Smirnov statistic of 0.00997. The o-Ps delay distribution similarly agreed with the configured model with $\tau = 3.0$ ns, yielding a Kolmogorov–Smirnov statistic of 0.01351. In both cases, the expected class and photon multiplicity were recovered for all events.

Equivalent fixed configuration and `FixedPsSourceEnvironmentProvider` runs produced byte-for-byte identical annihilation-summary and photon outputs. Disabling PsSource-specific truth recording also left the generated photon physics unchanged, with all 200 compared photon rows identical between the truth-enabled and truth-disabled runs. The complete transport-coupled regression results are summarized in **Table I**. These tests demonstrate that the annihilation result depends on the resolved environment and physics configuration, but not on whether the environment is supplied directly or through the provider interface, nor on whether optional PsSource truth output is enabled.

**Table I** — Transport-coupled regression results

| Case | N | Class/topology | Delay | Max Position residual | Max Timing residual | Statistical/equivalence Result |
|---|---|---|---|---|---|---|
| **Direct 2γ** | 100 | 100/100 class 0; 2γ/event | 0 ns fixed | $<1.0 \times 10^{-6}$ mm | 0 ns | PASS |
| **p-Ps 2γ** | 100 | 100/100 class 1; 2γ/event | 3 ns fixed | $<1.0 \times 10^{-6}$ mm | 0 ns | PASS |
| **o-Ps 2γ** | 100 | 100/100 class 2; 2γ/event | 3 ns fixed | $<1.0 \times 10^{-6}$ mm | 0 ns | PASS |
| **o-Ps 3γ** | 100 | 100/100 class 3; 3γ/event | 3 ns fixed | $<1.0 \times 10^{-6}$ mm | 0 ns | PASS |
| **p-Ps 2γ** | 10,000 | 10,000/10,000 class 1; 2γ/event | exponential, $\tau = 0.125$ ns | $<1.0 \times 10^{-6}$ mm | $1.0 \times 10^{-6}$ ns | KS = 0.00997; limit = 0.0163; PASS |
| **o-Ps 3γ** | 10,000 | 10,000/10,000 class 3; 3γ/event | exponential, $\tau = 3.0$ ns | $<1.0 \times 10^{-6}$ mm | $1.0 \times 10^{-6}$ ns | KS = 0.01351; limit = 0.0163; PASS |
| **Fixed vs provider** | 100 + 100 | Identical summary and photon outputs | — | — | — | Byte-for-byte identical; PASS |
| **Truth vs no truth** | 100 + 100 | Identical photon-physics rows | — | — | — | 200/200 photon rows identical; PASS |

### 3.2 External integration and independence

The reusable PsSource library was successfully integrated into standalone Geant4 host applications without requiring the legacy explicit photon generator or PsSource-specific event infrastructure. In these applications, the host retained responsibility for positron creation, geometry, materials, electromagnetic transport, and run control. PsSource was added only through registration of `PsSourceAnnihilationPhysics`, after which the existing terminal positron process was replaced and the generated annihilation photons continued through ordinary Geant4 tracking.

CTest integration applications completed successfully for the basic transport-coupled configuration, the fixed environment-provider configuration, and a host-defined provider that selected the local positronium environment from the transported terminal-state geometry context. The fixed fallback environment and the equivalent `FixedPsSourceEnvironmentProvider` configuration produced identical deterministic outputs, while the local provider demonstrated that environment selection could depend on terminal-state information without modification of the annihilation process.

Transport-coupled operation also remained independent of PsSource-specific truth and output classes. The standalone integration and installed-consumer tests ran without requiring `PositroniumGenerator`, `PositroniumTruthInfo`, a PsSource event action, a PsSource tracking action, or a prescribed output format. When truth recording was disabled, the generated photon properties were unchanged, confirming that observation and provenance reporting were not coupled to branch selection, delay sampling, or final-state generation.

The library was installed as a CMake package and consumed by a separate downstream project using `find_package(PsSource REQUIRED)` and the exported target `PsSource::PsSource`. The downstream project configured, compiled, linked, and executed successfully using only the installed library and public headers. Together, these tests demonstrated that PsSource could be incorporated as a terminal-annihilation physics extension while leaving the host source, geometry, detector, scoring, and downstream transport interfaces under host control.

### 3.3 Independent analytic validation of the Ore–Powell backend

The PsSource Ore–Powell backend reproduced the independently evaluated analytic three-photon energy distribution across the allowed two-dimensional phase space. The benchmark included 100,000 accepted o-Ps 3γ events, corresponding to 300,000 annihilation photons. No multiplicity failures or phase-space violations were identified.

The sampled joint-energy density closely matched the analytic Ore–Powell reference throughout the physical region shown in **Figure 1A** and **Figure 1B**. The standardized-residual map in **Figure 1C** was centered near zero and showed no coherent spatial structure indicative of a systematic sampling bias. The largest absolute standardized residual was 3.43.

For the 1,275 populated bins included in the global comparison, the chi-square statistic was

$$\chi^2 = 1236.91$$

with 1,274 degrees of freedom. The corresponding standardized chi-square score was

$$z_{\chi^2} = -0.735$$

which was consistent with finite-sample statistical variation. The maximum difference between the sampled and analytic marginal cumulative-distribution functions was 0.00208, substantially below the predefined tolerance of 0.01897.

Independent event-level consistency checks also passed. The maximum direction-vector normalization error was $8.31 \times 10^{-13}$, the maximum normalized energy-sum error was $1.57 \times 10^{-7}$, and the maximum opening-angle cosine error was $4.15 \times 10^{-10}$. All predefined analytic and event-level validation criteria were satisfied.

These results demonstrate that the PsSource implementation reproduces the QED-weighted Ore–Powell distribution rather than merely satisfying photon multiplicity and energy–momentum conservation. The analytic density, sampled density, and standardized residuals are shown in **Figure 1**.

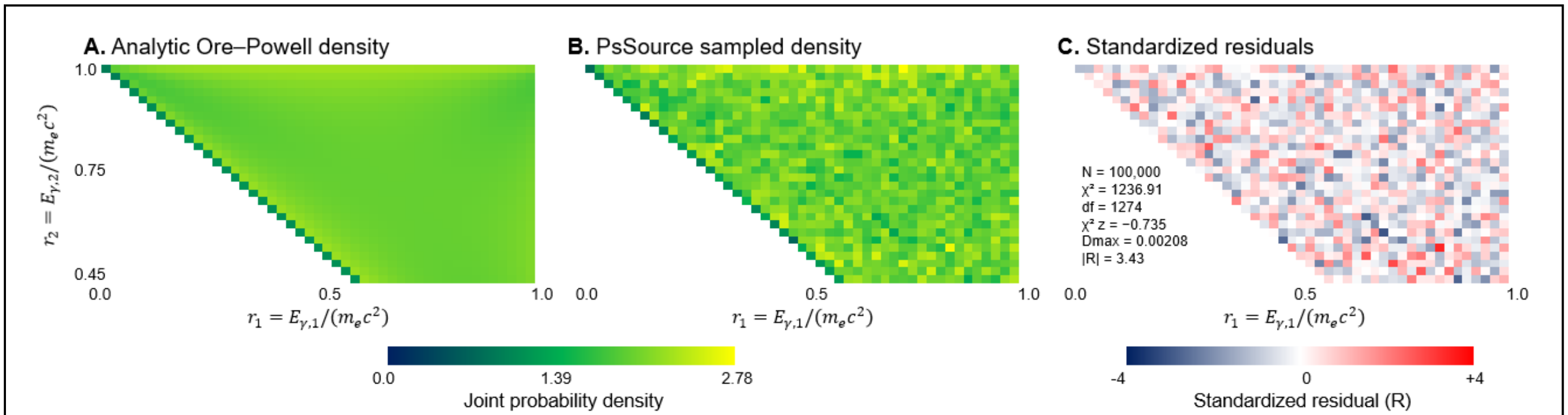

**Figure 1. Independent analytic validation of the PsSource Ore–Powell three-photon energy distribution.** (A) Analytic Ore–Powell joint probability density over the physically allowed two-dimensional energy phase space, expressed using the normalized photon energies $r_1 = E_{\gamma,1}/(m_e c^2)$ and $r_2 = E_{\gamma,2}/(m_e c^2)$. (B) Corresponding joint density sampled from 100,000 o-Ps 3γ events generated with the PsSource Ore–Powell backend. (C) Standardized residuals between sampled and expected bin counts. The comparison yielded $\chi^2 = 1236.91$ for 1274 degrees of freedom, $z_{\chi^2} = -0.735$, a maximum marginal cumulative-distribution difference of $\mathrm{D}_{max} = 0.00208$, and a maximum absolute standardized residual of $max_j |R_j| = 3.43$. The absence of coherent structure in the residual map indicates agreement with the analytic Ore–Powell distribution across the allowed phase space.

### 3.4 Agreement with native Geant4 and GATE

The ordered photon-energy distributions generated by PsSource agreed closely with the corresponding native Geant4 and GATE reference distributions. For each accepted o-Ps 3γ event, the photon energies were normalized to $m_e c^2$ and ordered from lowest to highest energy. The resulting lowest-, middle-, and highest-energy spectra are shown in **Figure 2A-C**.

Across all three ordered photon ranks, the PsSource, native Geant4, and GATE probability-density curves were visually superimposed. The expected rank-dependent structure was reproduced consistently: the lowest-energy photon occupied the lower portion of the allowed energy range, the middle-energy photon formed an intermediate distribution, and the highest-energy photon approached the upper kinematic boundary.

The pairwise empirical-CDF differences are shown in **Figure 2D**. Comparisons were performed for PsSource versus native Geant4, PsSource versus GATE, and native Geant4 versus GATE. The maximum absolute pairwise differences were approximately 0.004–0.005 across the ordered photon distributions. All ordered-rank empirical-CDF differences were below 0.00452 and showed no systematic rank-dependent discrepancy among the implementations.

The agreement was distributional rather than event-by-event because the three datasets were generated independently and did not share matched random-number sequences. Nevertheless, no systematic displacement, broadening, or rank-dependent distortion of the PsSource spectra was observed relative to either reference implementation.

These results confirm that PsSource correctly invokes the native Geant4 Ore–Powell model, extracts and orders the generated photon energies without modification, and reproduces the corresponding source-level distributions obtained through native Geant4 and GATE execution pathways. Agreement was expected because all three implementations ultimately use Geant4 Ore–Powell physics; the comparison therefore verifies implementation, data conversion, and cross-framework interoperability rather than providing an independent validation of the underlying physical model. Independent physical validation was provided by the analytic comparison in Section 3.3.

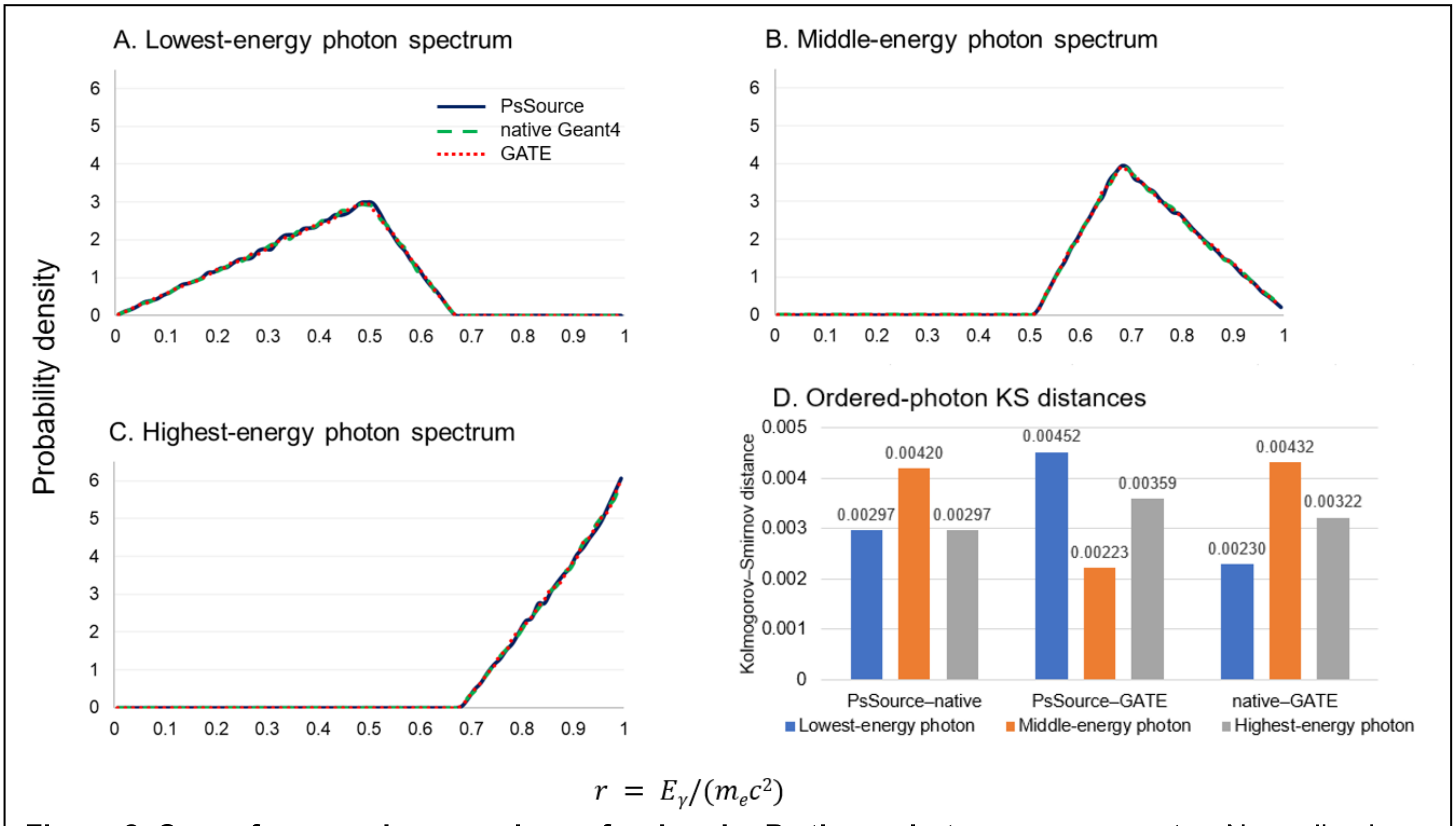

**Figure 2. Cross-framework comparison of ordered o-Ps three-photon energy spectra.** Normalized photon-energy distributions are shown for the lowest-energy (A), middle-energy (B), and highest-energy (C) photons from PsSource, native Geant4, and GATE. For each event, photon energies were normalized as $r = E_\gamma / m_e c^2$ and sorted in ascending order before construction of the marginal probability-density distributions. The three implementations produced nearly superimposed spectra across all ordered photon ranks. (D) Kolmogorov–Smirnov distances calculated separately for the lowest-, middle-, and highest-energy photon distributions for each framework pair. The maximum ordered-rank KS distances were 0.00420 for PsSource versus native Geant4, 0.00452 for PsSource versus GATE, and 0.00432 for native Geant4 versus GATE.

### 3.5 Rotational isotropy and coordinate-axis symmetry

The PsSource Ore–Powell backend produced isotropic photon directions and event-plane orientations. The analysis included 100,000 o-Ps 3γ events, corresponding to 300,000 annihilation photons.

The photon polar-angle distribution was uniform in $\cos\theta$, and the azimuthal-angle distribution was uniform in $\phi$, as expected for an isotropically oriented source. The corresponding goodness-of-fit tests passed for both angular variables. The photon $\cos\theta$ distribution yielded $\chi^2 = 6.74$ for 19 degrees of freedom, and the photon $\phi$ distribution yielded $\chi^2 = 20.78$ for 23 degrees of freedom. No localized excess or deficit associated with a preferred simulation axis was visible in either distribution.

The three-photon decay-plane orientation was independently assessed using the normal vector formed from two photon directions. Both the event-plane-normal $|cos\ \theta|$ and $\phi$ distributions were consistent with their uniform isotropic expectations. The event-plane-normal $|cos\ \theta|$ distribution yielded $\chi^2 = 17.70$ for 19 degrees of freedom, and the event-plane-normal $\phi$ distribution yielded $\chi 2 = 16.82$ for 23 degrees of freedom. Together, the photon-level and event-plane-level results showed that the event generator did not favor a particular polar or azimuthal direction.

The Cartesian-component analysis provided a direct test for x-, y-, or z-axis preference. The mean absolute components

$$\langle |k_x| \rangle, \qquad \langle |k_y| \rangle, \qquad \langle |k_z| \rangle$$

were each close to the isotropic expectation of 0.5. The mean absolute Cartesian components were 0.49987, 0.49991, and 0.50011 for the x, y, and z directions, respectively. The maximum difference between any pair of Cartesian means was

$$\Delta_{\text{axis}} = 2.43 \times 10^{-4}$$

providing no evidence of residual axis alignment.

All photon-direction, event-plane-normal, and Cartesian-symmetry checks satisfied their predefined criteria. These results provide no evidence that the generation, extraction, or event-construction pathway introduced measurable global coordinate-axis bias. The angular distributions and Cartesian summary metrics are shown in **Figure 3**.

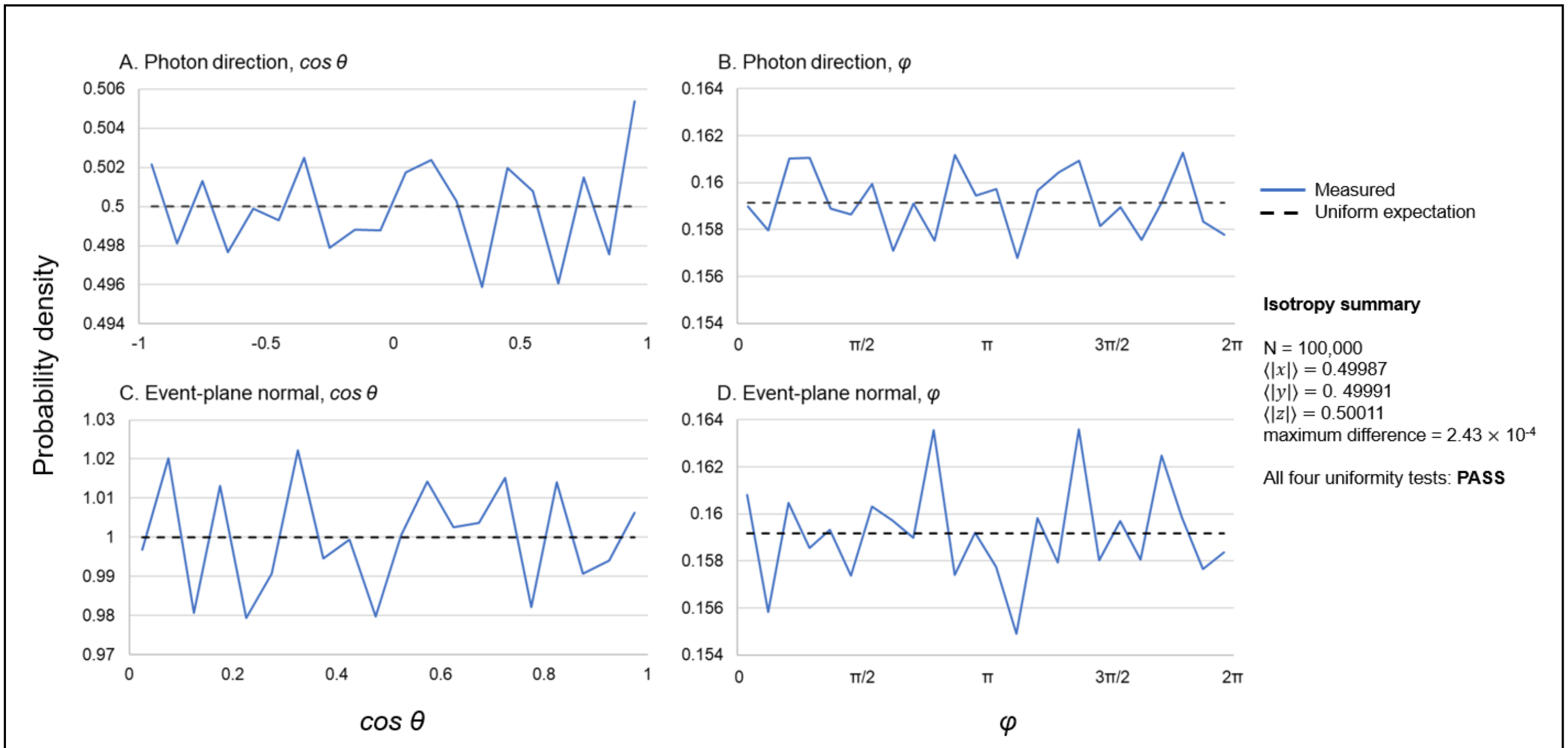

**Figure 3. Rotational isotropy and Cartesian-axis symmetry of PsSource Ore–Powell events.** Probability-density distributions are shown for the photon direction in $cos\ \theta$ (A) and $\phi$ (B), and for the three-photon event-plane normal in $|cos\ \theta_n|$ (C) and $\phi n$ (D). Solid blue lines show the measured distributions from 100,000 o-Ps 3γ events, and dashed black lines show the corresponding uniform expectations. The event-plane inclination is expressed as $|cos\ \theta_n|$ because the normals $\hat{n}$ and $-\hat{n}$ describe the same unoriented decay plane. The mean absolute photon-direction components were $\langle|k_x|\rangle = 0.49987,\ \langle|k_y|\rangle = 0.49991,\ \langle|k_z|\rangle = 0.50011$ with a maximum pairwise difference of $2.43 \times 10^{-4}$. All four angular-uniformity tests returned PASS, providing no evidence of photon-direction, event-plane, or Cartesian-axis bias.

### 3.6 Polarization-vector and symmetry validation

The polarized Ore–Powell backend generated valid polarization information for all 300,000 annihilation photons in the 100,000-event validation sample. The analysis evaluated vector normalization, direction–polarization transversality, polarization structure relative to the three-photon decay plane, global-coordinate symmetry, and preservation of the underlying Ore–Powell energy distribution.

#### 3.6.1 Normalization and transversality

Photon propagation directions and polarization vectors remained normalized to numerical precision. The maximum direction-normalization error over the full sample was

$$\delta_k^{max} = 8.39 \times 10^{-13}$$

and the maximum polarization-normalization error was

$$\delta_\varepsilon^{max} = 8.35 \times 10^{-13}$$

The polarization vectors were also transverse to their corresponding photon directions. The maximum absolute direction–polarization scalar product was

$$\delta_\perp^{max} = 1.50 \times 10^{-12}$$

All three maximum errors were many orders of magnitude below the predefined validation tolerance,

$$\delta_{tol} = 2 \times 10^{-5}$$

and no photon lacked a valid polarization assignment. The normalization and transversality checks therefore returned an overall status of PASS.

These results show that polarization vectors were propagated through the PsSource event-construction and truth-output pathways without detectable loss of normalization or orthogonality. The corresponding summary values are reported in **Figure 4A**.

3.6.2 Decay-plane structure and global-axis isotropy

The polarization vectors exhibited a clear relationship to the three-photon decay plane. The absolute out-of-plane projection, $P_{\perp} = |\varepsilon \cdot \hat{n}|$, was distributed preferentially toward larger values, with a sample mean of 0.7011. The corresponding in-plane component, $P_{\parallel}$, showed the complementary behavior expected from the unit normalization of the polarization vector. These distributions demonstrate a reproducible relationship between the generated polarization vectors and the three-photon decay plane.

Global-coordinate symmetry was assessed using the absolute polarization projection onto the simulation z axis,

$$P_z = |\varepsilon \cdot \hat{z}|$$

For an isotropically oriented population of polarization vectors, the expected mean is

$$\langle P_z \rangle = 0.5$$

The observed value was

$$\langle P_z \rangle = 0.49960$$

which was consistent with the isotropic expectation. The distribution of the absolute global-$z$ polarization projection, $|\varepsilon \cdot \hat{z}|$, was consistent with uniformity ($\chi^2 = 31.00$, 49 degrees of freedom; standardized $\chi 2\ score\ z = -1.82$), indicating no detectable preference for the global-$z$ direction. Together, these results show that the polarized backend produces reproducible decay-plane-dependent polarization structure while avoiding spurious alignment with the simulation axes. The decay-plane projections and global-axis isotropy results are shown in **Figure 4B and Figure 4C**.

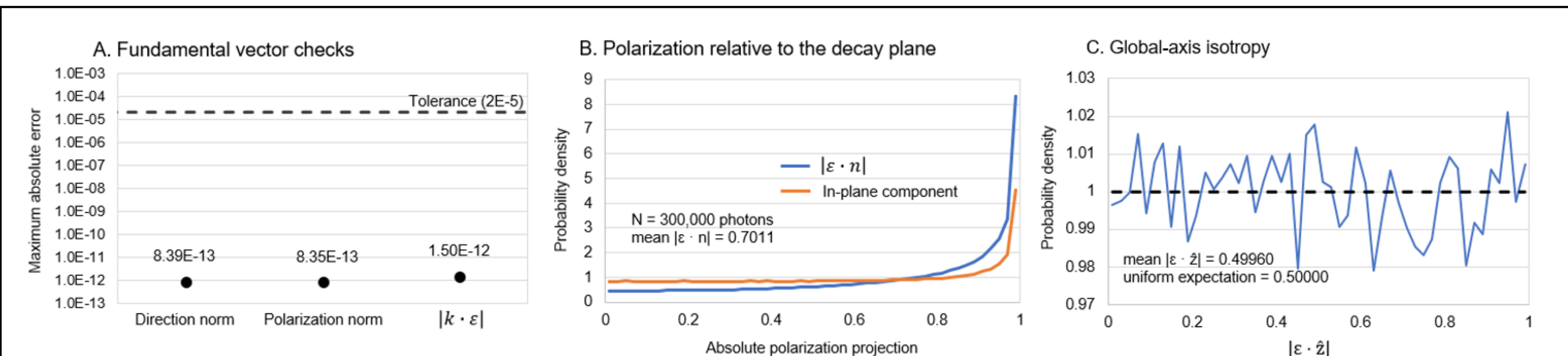

**Figure 4. Validation of the polarized Ore–Powell backend.**
(A) Maximum direction-normalization, polarization-normalization, and direction–polarization transversality errors across 300,000 annihilation photons. All values were no greater than $1.50\times10^{-12}$, well below the predefined tolerance of $2\times10^{-5}$. (B) Polarization relative to the three-photon decay plane, represented by the absolute out-of-plane projection $|\ \varepsilon \cdot \hat{n}\ |$ and the complementary in-plane component. The mean absolute out-of-plane projection was 0.7011. (C) Distribution of the absolute polarization projection onto the global $z$ axis, $|\varepsilon\ \cdot\ \hat{z}|$. The measured mean was 0.49960, and the distribution was consistent with uniformity ($\chi 2 = 31.00$ , 49 degrees of freedom; standardized chi-square score $z\chi^2 = -1.82$), indicating no detectable global-axis preference.

The polarized and ordinary Ore–Powell samples showed a maximum empirical-CDF difference of 0.00168, below the predefined acceptance threshold of 0.020. Enabling polarization therefore did not measurably alter the underlying photon-energy distribution.

The present validation was limited to photon-level polarization properties because the parent o-Ps magnetic substate was not retained. Magnetic-substate-resolved observables were therefore not evaluated.

**4. Discussion**

The principal contribution of this work is the development and validation of PsSource as an installable Geant4 extension for configurable positronium-aware terminal annihilation. PsSource acts only when a transported positron reaches the terminal at-rest annihilation stage, preserving the host application's source definition, positron transport, geometry, materials, detector construction, scoring, digitization, reconstruction, and output infrastructure. The results establish two complementary aspects of the implementation. First, the transport-coupled integration behaves as intended: the transported terminal position and global time are preserved, the positronium delay is applied after transport, the configured annihilation class is realized correctly, and the resulting photons are returned as ordinary Geant4 secondaries. Second, the supported three-photon final states reproduce the expected Ore–Powell energy, angular, and polarization behavior. By separating host-supplied environmental parameters from annihilation final-state generation, PsSource provides a reusable interface for fixed or spatially varying positronium models without embedding assumptions about tissue- or material-specific behavior in the core physics implementation.

The transport-coupled boundary is central to the practical value of PsSource because it preserves the physical history already established by the host simulation. A source-level photon generator defines the annihilation position, time, and final state before Geant4 has transported the positron, thereby bypassing the simulated positron range, accumulated transport time, terminal geometry, and local material context. PsSource instead intervenes only at terminal at-rest annihilation, using the transported positron endpoint as the annihilation position, adding the configured positronium delay to the terminal global time, and returning the resulting photons as ordinary Geant4 secondaries. This allows positronium-aware annihilation to be introduced without replacing the host application's transport model or downstream detector response. The boundary is deliberately limited to terminal at-rest annihilation and does not modify in-flight annihilation processes.

Separating local environment resolution from annihilation final-state generation allows PsSource to support heterogeneous applications without embedding material- or tissue-specific assumptions in the core physics process. A fixed environment can be used for homogeneous simulations and controlled studies, whereas a host-defined provider can assign parameters from terminal-state information such as material, region, volume, position, or an external spatial model. This design also provides a pathway for future experimentally derived or tissue-specific parameterizations to be incorporated without changing the annihilation implementation itself. PsSource verifies that the supplied fractions, lifetimes, and branch definitions are finite, non-negative, and internally consistent, but these checks establish numerical validity rather than physical realism. The host application or separately validated environment model remains responsible for ensuring that the selected lifetimes and annihilation fractions are appropriate for the local medium, because these quantities are environment dependent and should not be interpreted as universal or freely independent material properties.

The transport-coupled regression results demonstrate that the architectural requirements of PsSource are realized within actual Geant4 transport rather than only within isolated source-generation tests. Across the four supported annihilation classes, the transported terminal position and global time were preserved, the final annihilation time decomposed correctly into positron transport time plus the sampled positronium delay, and the expected photon multiplicity, class identity, creator-process provenance, and positron parentage were recovered. Equivalent fixed and provider-supplied environments produced identical outputs under matched random conditions, confirming that environment resolution can be exchanged without altering the underlying physics. Likewise, disabling PsSource-specific truth recording left the generated photon states unchanged, demonstrating that observation and provenance are optional interfaces rather than components of the annihilation mechanism. The fixed- and exponential-delay tests therefore verify correct implementation, propagation, and statistical sampling of configured lifetimes; they do not establish that any selected lifetime or branch fraction is biologically or materially valid.

The independent analytic comparison provides the strongest validation of the three-photon final-state physics because photon multiplicity, energy conservation, and momentum closure alone cannot distinguish a physical sampler from an incorrect distribution that satisfies the same constraints. The Ore–Powell events generated through the PsSource interface reproduced the independently evaluated two-dimensional QED-weighted energy distribution, with $\chi^2 = 1236.91$ for 1274 degrees of freedom, a standardized $\chi^2$ score of -0.735, and a maximum marginal empirical-CDF difference of 0.00208. This agreement indicates that the native Geant4 Ore–Powell final state was correctly invoked, extracted, and transferred through the PsSource pathway without measurable distortion of the underlying three-photon energy distribution.

The cross-framework comparisons provide an additional check that the PsSource integration preserves the energy distributions produced by established Ore–Powell pathways. PsSource delegates three-photon generation to the native Geant4 model, so the PsSource and native Geant4 datasets should not be interpreted as independent physical models. They were, however, generated and processed through separate execution and analysis pathways, making agreement between them useful for detecting integration, extraction, or data-handling errors. The GATE dataset provides a more distinct framework-level comparison, and the close agreement of the ordered photon-energy spectra across all three pathways supports correct transfer of the Ore–Powell final state through PsSource. Because the comparisons used different Geant4 versions and independently executed workflows, the results demonstrate distributional consistency rather than binary or event-by-event identity.

The rotational-isotropy and coordinate-axis tests address a potential source of systematic error that would not necessarily be revealed by energy-spectrum agreement alone. Photon directions and three-photon event-plane normals were consistent with their isotropic expectations, and the mean absolute Cartesian direction components agreed closely across the (x), (y), and (z) axes. These results provide no evidence that the complete generation, extraction, conversion, or analysis pathway introduced a preferred simulation axis. Avoiding such artifacts is important because even a small coordinate-dependent bias could propagate into detector-acceptance estimates, angular-correlation measurements, or polarization-sensitive analyses and be misinterpreted as a physical effect.

The polarized pathway produced normalized photon directions and polarization vectors that remained transverse to their associated propagation directions to numerical precision. The ensemble also showed reproducible decay-plane-dependent polarization structure without evidence of preferred alignment with a global simulation axis. Comparison with the ordinary Ore–Powell sample further showed no detectable alteration of the underlying photon-energy distribution, indicating that polarization information was added without distorting the established three-photon kinematics. These findings are limited to photon-level ensemble characterization. PsSource does not retain the parent ortho-positronium magnetic substate, and the present validation therefore does not support magnetic-substate-resolved interpretations or detector-level claims regarding polarization sensitivity.

Several limitations define the scope of the present implementation and validation. The terminal at-rest handoff is a computational integration boundary rather than a microscopic model of positronium formation. PsSource does not simulate positronium formation during positron slowing, positronium-atom transport or diffusion, collision-by-collision pickoff, quenching, or spin conversion, and it does not infer material- or tissue-specific branch fractions or lifetimes automatically. The current implementation also does not include two-photon polarization or entanglement modeling or magnetic-substate-resolved positronium physics. The annihilation-physics benchmarks were performed through the explicit validation pathway rather than within a complete detector simulation, so the present results do not establish detector sensitivity, image quality, or reconstruction performance. Validation was performed principally with Geant4 11.3.2, while the GATE reference used Geant4 11.3.0; broad equivalence across Geant4 versions and formal multithreaded equivalence were not evaluated.

PsSource is intended to complement rather than supersede native Geant4, GATE, or detector-specific simulation frameworks. Its value lies in providing a reusable process-level integration layer in which host applications continue to create and transport positrons normally, local positronium parameterizations can be replaced without modifying the annihilation process, and validated two- and three-photon final states are returned as ordinary Geant4 secondaries. Optional observation and provenance interfaces can be enabled when needed but are not required for physics execution, and the installable library can be adopted without using the PsSource benchmark applications, command-line tools, or output schema. Future work should focus on validated material- and tissue-specific environment models, explicit positronium transport and diffusion, magnetic-substate-resolved polarization, broader Geant4-version and multithreaded validation, and detector-level studies of positronium-sensitive imaging, spectroscopy, and fundamental annihilation measurements.

## 5. Conclusion

PsSource 3.0 provides an installable Geant4 extension for configurable positronium-aware terminal annihilation. The transport-coupled architecture preserves the host application's positron source, transport history, terminal position and time, geometry, detector response, scoring, and output systems while replacing only terminal at-rest annihilation. Fixed or host-defined local environments determine the realized annihilation class and delay, and the resulting two- or three-photon final states are returned as ordinary Geant4 secondaries. Regression testing confirmed preservation of terminal-state information, correct timing decomposition, recovery of all supported annihilation classes, proper photon multiplicity and parentage, equivalence of fixed and provider-supplied environments, and invariance of photon physics when PsSource-specific observation is disabled. Independent analytic, cross-framework, isotropy, and polarization tests further showed that the integrated Ore–Powell pathways preserve the expected three-photon energy, angular, and photon-level polarization behavior. PsSource therefore provides a validated process-level interface for incorporating configurable positronium annihilation into existing Geant4 applications while maintaining a clear separation between transport, local environmental parameterization, annihilation physics, and optional provenance.

## 6. Declarations

*Ethical Approval and Consent to participate*

This study involved software development, simulation, and computational validation only. No human participants, human data, human tissue, or animals were involved. Ethical approval and consent to participate were therefore not required.

*Consent for publication*

The author consents to publication of this manuscript.

*Availability of supporting data*

The PsSource source code, build and installation files, validation scripts, reference implementations, benchmark configurations, and processed figure data used in this study are available at https://github.com/parkerjg/PsSource. The software release evaluated in this manuscript is PsSource version 3.0.0, tag pssource-v3.0.0, release commit 7dd7fdc. The repository includes the installable CMake package, public transport-coupled integration interface, environment-provider and observer examples, transport regression workflows, analytic Ore–Powell validation, native Geant4 and GATE reference configurations, cross-framework comparisons, isotropy and three-photon geometry tests, polarization validation, and the processed data used to generate the reported figures and tables. The validated Geant4 baseline for the PsSource release was version 11.3.2.

*Competing interests*

The authors declare that they have no competing interests.

*Funding*

No specific funding was received for this work.

*Authors' contributions*

JGP conceived the study, developed the software, designed the validation experiments, performed the simulations and analyses, interpreted the results, and wrote and revised the manuscript.

*Acknowledgements*

The authors acknowledge the Indiana University Pervasive Technology Institute for providing supercomputing and storage resources that have contributed to the research results reported within this paper (https://pti.iu.edu/).

*Authors' information*

Jason G. Parker, PhD, Department of Radiology and Imaging Sciences, Indiana University School of Medicine, Indianapolis, Indiana, USA. Corresponding author email: parkerjg@iu.edu

## 7. References

1. Conti, M. and L. Eriksson, *Physics of pure and non-pure positron emitters for PET: a review and a discussion.* EJNMMI Phys, 2016. 3(1): p. 8.
2. Kaminska, D., et al., *A feasibility study of ortho-positronium decays measurement with the J-PET scanner based on plastic scintillators.* Eur Phys J C Part Fields, 2016. 76(8): p. 445.
3. Moskal, P., et al., *Feasibility study of the positronium imaging with the J-PET tomograph.* Phys Med Biol, 2019. 64(5): p. 055017.
4. Moskal, P., et al., *Positronium imaging with the novel multiphoton PET scanner.* Sci Adv, 2021. 7(42): p. eabh4394.
5. Moskal, P., et al., *Positronium image of the human brain in vivo.* Sci Adv, 2024. 10(37): p. eadp2840.
6. Moskal, P., et al., *Positronium in medicine and biology.* Nature Reviews Physics, 2019. 1(9): p. 527–529.
7. Ore, A. and J.L. Powell, *Three-Photon Annihilation of an Electron-Positron Pair.* Physical Review, 1949. 75:1696-99.
8. Qi J, Huang B. Positronium lifetime image reconstruction for TOF PET. *IEEE Transactions on Medical Imaging.* 2022;41(10):2848–2855. doi:10.1109/TMI.2022.3174561.
9. Mercolli L, Steinberger WM, Läppchen T, et al. In vivo voxel-wise positronium lifetime imaging of thyroid cancer using clinically routine I-124 PET/CT. *EANM Innovation.* 2026;2:100017. doi:10.1016/j.eanmi.2025.100017.
10. Agostinelli, S., et al., *GEANT4-a simulation toolkit.* Nuclear Instruments & Methods in Physics Research Section a-Accelerators Spectrometers Detectors and Associated Equipment, 2003. 506(3): p. 250–303.
11. Allison, J., et al., *Recent developments in GEANT4.* Nuclear Instruments & Methods in Physics Research Section a-Accelerators Spectrometers Detectors and Associated Equipment, 2016. 835: p. 186–225.
12. Collaboration, G. *Physics Reference Manual*. 11.3 (doc Rev9.2) 2024; Available from: https://geant4-userdoc.web.cern.ch/UsersGuides/PhysicsReferenceManual/BackupVersions/V11.3b/html/index.html.
13. Sarrut, D., et al., *The OpenGATE ecosystem for Monte Carlo simulation in medical physics.* Phys Med Biol, 2022. 67(18).
14. Salvadori, J., et al., *PET digitization chain for Monte Carlo simulation in GATE.* Phys Med Biol, 2024. 69(16).
15. Moskal, P., et al., *Feasibility studies of the polarization of photons beyond the optical wavelength regime with the J-PET detector.* Eur Phys J C Part Fields, 2018. 78(11): p. 970.
16. Tashima H, Takyu S, Nishikido F, Takahashi M, Yamaya T. Modeling positronium lifetime distribution in Geant4 Monte Carlo simulation. In: *2023 IEEE Nuclear Science Symposium, Medical Imaging Conference and Room Temperature Semiconductor Detector Conference (NSS MIC RTSD).* 2023.